\documentclass{WileyMSP-template}

\usepackage{amsmath}
\usepackage{dsfont}
\usepackage{nicefrac}

\begin{document}

\pagestyle{fancy}
\rhead{\includegraphics[width=2.5cm]{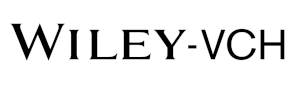}}

\title{Field dynamics inference for local and causal interactions}

\maketitle

% Author: Please give full first and last names for authors and include * after the name of all corresponding authors

\author{Philipp Frank*}
\author{Reimar Leike}
\author{Torsten A. En\ss lin}

% Dedication

%\dedication{Optional dedication here. If no dedication is required, please leave blank}

% Affiliations: Please provide adacemic titles (Prof. or Dr.) for all authors where applicable, and include an institutional email address for all corresponding authors
\begin{affiliations}
Ph. Frank, R. Leike, PD.\ Dr.\ T. A. En\ss lin\\
Max-Planck Institut f{\"u}r Astrophysik, Karl-Schwarzschild-Stra\ss e 1, 85748, Garching, Germany \\
Ludwig-Maximilians-Universit{\"a}t M{\"u}nchen, Geschwister-Scholl-Platz 1, 80539, M{\"u}nchen, Germany \\
philipp@mpa-garching.mpg.de
\end{affiliations}

% Keywords: Please provide a minimum of three and a maximum of seven keywords, separated by commas

\keywords{Information theory, Stochastic processes, Data analysis, Bayesian methods}

% Abstract should be written in the present tense and impersonal style (i.e., avoid we), and be at most 200 words long
\begin{abstract}
Inference of fields defined in space and time from observational data is a core discipline in many scientific areas. This work approaches the problem in a Bayesian framework. The proposed method is based on statistically homogeneous random fields defined in space and time and demonstrates how to reconstruct the field together with its prior correlation structure from data. The prior model of the correlation structure is described in a non-parametric fashion and solely builds on fundamental physical assumptions such as space-time homogeneity, locality, and causality. These assumptions are sufficient to successfully infer the field and its prior correlation structure from noisy and incomplete data of a single realization of the process as demonstrated via multiple numerical examples.
\end{abstract}

% Text: Please use section headings and subheadings as specified below. For communications, all section headings apart from Experimental Section should be removed
% Please make the first reference to a display item bold: \textbf{Figure 1}
% Do not abbreviate Figure, Equation, etc.; display items are always singular, i.e., Figure 1 and 2.
% Equations are always singular, i.e., Equation 1 and 2, and should be inserted using the {equation} environment, not as graphics
% Please do not use footnotes in the text, additional information can be added to the Reference list.

\section{Introduction}
Modeling as well as inferring quantities defined in space and time on the basis of observational data thereof has always been at the very core of many scientific areas. In recent years, astrophysical imaging began to become sensitive to the temporal dimension, in addition to the spatial ones. This is due to the fact that although large astrophysical objects such as galaxies appear to be static on observational timescales, small objects such as stars and binary black holes exhibit transient, periodic, and quasi-periodic modulations of the emission on observable timescales.

In addition, the spatio-temporal correlation structure of non-astronomical systems plays a central role in the calibration of modern telescopes. In particular the temporal variability of these systems is used in order to identify and distinguish them from the typically static astronomical object of interest. As a prominent example, modern radio telescopes such as LOFAR \cite{2013A&A...556A...2V} or the upcoming SKA are limited in resolution by the deflection of incoming radio signals due to the ionosphere. The strength of these distortions ultimately depends on the electron density of the ionosphere. As this density is not known for all observed locations $x$ at time $t$, it has to be inferred along with the incoming flux. Typically, the electron density is probed via observing a calibration target with known flux at location $x'$ and time $t'$. Therefore, it is also necessary to make a statement about the correlation structure of the electron density in order to extrapolate the information gained at $(t',x')$ to the space-time location $(t,x)$ where the actual observation is made.

In order to tackle these as well as other inference problems of this kind in space and time, we have to perform inference of continuous quantities, or fields, from a finite set of measurement data. This problem is in general ill-posed, as we aim to constrain infinite degrees of freedoms (dofs) on finite, usually also noisy, measurements. Consequently we rely on Bayesian inference, more precisely on Information Field Theory (IFT) \cite{2019AnP...53100127E}, and use this language to encode prior knowledge about the system under consideration. Typically, this prior knowledge is incomplete, and there exist a set of unknown hyper-parameters, such as the spatio-temporal prior correlation structure, which have to be inferred along with the field of interest. We outline in this paper how physically motivated concepts such as spatio-temporal homogeneity, locality as well as causality can be encoded into the prior correlation structure. Furthermore, we demonstrate that the resulting hierarchical prior model is restrictive enough to perform inference on the basis of noisy and incomplete data of a single realization of the process, while still being flexible enough to capture complicated correlation structures.

Traditionally there exist two different pictures on random fields in space-time, one results in space and time being treated separately \cite{protter2005stochastic,grecksch1996time}, while the other models the field as defined over a single space, namely space-time \cite{doering1987stochastic,roberts2003step} (see e.g.\ \cite{arnold1974stochastic} for an extensive discussion). In this work we rely on the latter picture. Consequently the corresponding inference problems can be regarded as the task of inferring a field defined on a single space, given a finite amount of measurements in this space.

To this end, in section \ref{sec:ift} we start with a brief introduction to IFT and the notation used in this work. In section \ref{sec:prior} we discuss how to encode our prior knowledge about the field and its correlation structure into a joint prior distribution thereof. This prior is then used in section \ref{sec:inference} in order to solve the corresponding inference problem and the performance of the resulting algorithm is demonstrated in section \ref{sec:application}. Ultimately, in section \ref{sec:conclusion}, we conclude the paper with a brief summary of the proposed concepts.

\section{Information Field Theory and Gaussian processes}\label{sec:ift}
Information Field Theory is a statistical field theory that aims to describe Bayesian inference of fields defined over some continuous space, or space-time.
For simplicity, we first consider a one dimensional random process, and provide an extension to space-time at the end of this section. To this end, consider a zero mean square-integrable random process $s$ defined over a closed interval $I = [0,B]$, i.E.\ $s^x \in L^2(I)$ with probability $P(s)$. We define the covariance function $S$ as
\begin{equation}
	S(x,y) \equiv \left< s^x \left(s^y\right)^*\right>_{P(s)} \ ,
\end{equation}
where $^*$ denotes complex conjugation. If we associate with $S$ a linear operator $O_S : L^2(I) \rightarrow L^2(I)$ and define its application via
\begin{equation}
	(O_S s)^x \equiv \int_{I} S(x,y) s^y \ \mathrm{d}y \ ,
\end{equation}
we may define the eigenvalues $\lambda_k$ and eigenfunctions $e_k$ of the linear operator $O_S$ via
\begin{equation}
	(O_S \ e_k)^x = \int_{I} S(x,y) \ e_k^y \ \mathrm{d}y = \lambda_k \ e_k^x \ .
\end{equation}
Since the eigenfunctions of $O_S$ form an orthonormal basis and the random process $s$ lies within the span of $e_k$, the Karhunen-Lo\`{e}ve theorem \cite{karhunen1947lineare, loeve1977probability} states that $s$ may be represented in this basis as
\begin{equation}
	s^x = \sum_{k = - \infty}^{\infty} e^x_k \ \tilde{s}^k  \ ,
\end{equation}
with the modes $\tilde{s}^k$ defined via
\begin{equation}
	\tilde{s}^k \equiv \int_{I} \left(e^x_k\right)^* s^x \ \mathrm{d}x 
	 \quad \text{with} \quad \left<\tilde{s}^k\right> = 0 \ , \left<\tilde{s}^k \ \tilde{s}^q\right> = \delta_{k q} \ \lambda_k \quad \forall k,q \in \mathds{Z} \ .
\end{equation}
Specifically all $\tilde{s}^k$ become zero mean and uncorrelated random variables with variance $\lambda_k$. Consequently, an inference problem of $s^x$ can be reduced to an inference problem in $\tilde{s}^k$

\subsection{Statistically homogeneous Gaussian processes}
Statistically homogeneous Gaussian processes are a special, but very useful, process for prior modeling of physical processes as the statistical homogeneity implies that a priori no specific location in $I$ is singled out. We may again define a zero mean random process $s^x$ with the additional requirement that the covariance takes the form
\begin{equation}
	S(x,y) = S(x-y) \ \forall x,y \in I \ .
\end{equation}
If we additionally require the space to obey periodic boundary conditions such that
\begin{equation}
	s^{x+B} = s^x \quad \text{and} \quad S(x+B,y) = S(x,y) \ ,
\end{equation}
the Wiener-Khinchin theorem \cite{wiener_extrapolation_1950} implies that the eigenbasis of the linear operator associated with this covariance function is the Fourier basis and its spectrum is the Fourier power spectrum. This allows for a representation of $s$ as
\begin{equation}\label{eq:mode_expansion}
	s^x = \sum_{k = - \infty}^{\infty} e^{\frac{2 \pi}{B} i k x} \ g^k \xi^k \quad \text{with} \quad \left|g^k\right|^2 = \lambda_k \ , \ \xi^k \sim \mathcal{G}(\xi^k, 1) \ \forall k \in \mathds{Z} \ ,
\end{equation}
where $i$ denotes the imaginary unit and $\xi^k$ are independent and identically distributed Gaussian random variables with mean zero and variance one. For a compact notation we may define the Fourier transformation $\mathcal{F}^k_x = e^{\frac{2 \pi}{B} i k x}$ and an infinite dimensional diagonal matrix $\widehat{G}^k_q = \delta_{k q} g^k$ to write
\begin{equation}\label{eq:fourier_basis}
	s^x = (\mathcal{F} \widehat{G} \xi)^x \equiv \sum_{k = - \infty}^\infty \sum_{q = - \infty}^\infty \mathcal{F}^x_k \widehat{G}^k_q \xi^q \ .
\end{equation}
The application of its adjoint, abbreviated as $\mathcal{F}^\dagger$, is defined as
\begin{equation}
	\tilde{s}^k = \left(\mathcal{F}^\dagger s\right)^k \equiv \int \left(\mathcal{F}^x_k\right)^* s^x \ \mathrm{d}x \ .
\end{equation}

\subsection{Linear Measurements and the Wiener Filter}\label{sec:wf}
In order to perform Bayesian inference of $s$ given some observational data $d$, we require a data generating model (or data-model) that describes how $d$ is obtained from $s$ and possibly additional nuisance parameters $n$ that describe the measurement noise. A simple but very powerful idealized model is a linear measurement of $s$ with additive Gaussian distributed noise $n$, independent of $s$, defined as
\begin{equation}\label{eq:data_model}
	d^i = (R s)^i + n^i = \int_{I} R^i_x s^x \ \mathrm{d}x \quad \forall i \in \left\lbrace 1, ..., M \right\rbrace \quad \text{with} \quad n \sim \mathcal{G}(n,N) \ ,
\end{equation}
where $R : L^2(I) \rightarrow \mathds{R}^M$ is a linear operator that maps onto a discrete $M$-dimensional space, called data space, and $\mathcal{G}(n,N)$ denotes a Gaussian distribution with zero mean and covariance $N$. A linear measurement operator may represent common scenarios such as measurements of values at single locations, integrated measurements over a specific area, partially masked areas, convolution with a point spread function, and linear combinations thereof.

We can represent the data-model (Eq.\ \eqref{eq:data_model}) as a generating process in terms of $\xi$ by inserting the Fourier basis representation of $s$ (Eq.\ \eqref{eq:fourier_basis}) to get
\begin{equation}
	d = R \mathcal{F} \widehat{G} \ \xi + n \ .
\end{equation}
The inference problem may thus be regarded as the task of constructing the posterior distribution of $\xi$, given $d$ and the background information i.E.\ the specific form of $R$ and the prior spectrum $g$ which in turn defines $\widehat{G}$. This problem allows for a closed form solution by means of quadratic completion (see e.g.\ \cite{2019AnP...53100127E}) and the posterior remains a Gaussian distribution with mean $m_\xi$ and covariance $D_\xi$ given as
\begin{gather}\label{eq:wf_xi}
\begin{aligned}
\begin{alignedat}[t]{3}
&D_\xi &&= \left(\widehat{G}^\dagger \mathcal{F}^\dagger R^\dagger N^{-1} R \mathcal{F} \widehat{G} + \mathds{1}\right)^{-1} &&= \mathds{1} - \widehat{G}^\dagger \mathcal{F}^\dagger R^\dagger \left(R \mathcal{F} \widehat{G} \widehat{G}^\dagger \mathcal{F}^\dagger R^\dagger + N \right)^{-1} R \mathcal{F} \widehat{G} \\
&m_\xi &&= D_\xi \widehat{G}^\dagger \mathcal{F}^\dagger R^\dagger N^{-1} d &&= \widehat{G}^\dagger \mathcal{F}^\dagger R^\dagger \left(R \mathcal{F} \widehat{G} \widehat{G}^\dagger \mathcal{F}^\dagger R^\dagger + N \right)^{-1} d
\end{alignedat}
\end{aligned}
\end{gather}

where the first part of the equations is the common representation of the Wiener Filter, now for a infinite number random variables $\xi$ which are the coefficients of the random process $s$ in the eigenbasis of its prior. The right hand side can be obtained by straightforward manipulation of the expressions. It has the convenient property that the only matrix inversion involved appears in the finite dimensional data-space and thus entirely avoids inversion of infinite dimensional matrices. 

The posterior of the coefficients $\xi$ can be used to construct the posterior of $s$ by insertion of the modes into the expansion of $s$ (Eq.\ \eqref{eq:fourier_basis}). Therefore the posterior mean $m$ and the covariance $D$ of $s$ are denoted as
\begin{equation}\label{eq:wf_s}
	m = \mathcal{F} \widehat{G} \ m_\xi \quad \text{and} \quad D = \mathcal{F} \widehat{G} D_\xi \widehat{G}^\dagger \mathcal{F}^\dagger \ .
\end{equation}
This concludes the description of the Wiener Filter theory applied to square integrable random processes in terms of the eigenbasis of the linear operator associated with the prior covariance. A mathematically more rigorous and coordinate free discussion of these concepts is beyond the scope of this work, but is described in great detail by e.g.\ \cite{stuart_2010}.

\subsection{Consistent discretization}\label{sec:discretization}
For many physically relevant choices of $R$ and $\widehat{G}$ the Fourier integrals involved in Eqs.\ \eqref{eq:wf_xi} and \eqref{eq:wf_s} may be difficult to solve or may not have a closed form representation analytically. Therefore, for practical applications, the inference problem is often discretized and the discrete problem is solved instead. However, as shown by e.g.\ \cite{1930-8337_2009_1_87}, care must be taken when defining a discretization in order to ensure that the finite dimensional approximation is consistent with the infinite dimensional inference problem. In this work, we achieve a discrete representation by truncating the Fourier series at a maximal / minimal value $\pm \nicefrac{K}{2}$. Specifically
\begin{equation}
	\bar{s}^x \equiv \sum_{k \in W} g^k \xi^k \ e^{\frac{2 \pi}{B} i k x} \quad \text{with} \quad W \equiv \left\lbrace-\nicefrac{K}{2}, ..., \nicefrac{K}{2} \right\rbrace\ .
\end{equation}
A measure for the discretization error may be defined by means of the expected squared difference between $\bar{s}$ and $s$ as
\begin{equation}
	\left(\epsilon^x\right)^2 \equiv \left< \left|s^x - \bar{s}^x\right|^2\right> = \sum_{k \in \mathds{Z} \setminus W} \left<\left|g^k \xi^k \ e^{\frac{2 \pi}{B} i k x}\right|^2\right> = \sum_{k \in \mathds{Z} \setminus W} \left|g^k\right|^2 \ \forall x \in I \ ,
\end{equation}
which quantifies the difference between the infinite dimensional process and the finite dimensional approximation of the quantity of interest $s$. In order to ensure that inference is consistent, the discretization error $\epsilon_R$ of the observed quantity $R s$ has to be considered. It is given as
\begin{equation}
	\left(\epsilon^j_R\right)^2 \equiv \left< \left|\left(R s\right)^j - \left(R \bar{s}\right)^j\right|^2\right> = \sum_{k \in \mathds{Z} \setminus W} \left|g^k\right|^2 \left|\left(R e_k\right)^i\right|^2 \quad \text{with} \quad e_k^x = e^{\frac{2 \pi}{B} i k x} \ .
\end{equation}
A small $\epsilon_R$ is sufficient to ensure that the discrete approximation of the inference problem is close to the continuous one as it ensures that the contribution of modes not in $W$ to the observed quantity is small and therefore the information gained about these modes via the observation is also small compared to the information gain about the modes in $W$. For posterior analysis of $s$ it is also relevant to have a good discrete approximation and therefore in general also $\epsilon$ should be small.

A minimal requirement is that the Gaussian process is continuous, which implies that $\left|g^k\right|^2$ decays asymptotically at least with $\nicefrac{1}{|k|^{2+\gamma}} \ , \gamma \geq 0$. This ensures that the series expansion of $s$ converges and that there exists a $K$ such that $\epsilon$ becomes small. Fortunately, the assumption of continuity is met by most physically relevant processes.

In general, the magnitude of $\epsilon_R$ can only be specified for a given measurement scenario as it depends on the specific form of the measurement operator $R$. However, as the properties of $R$ can fully be defined via its action on the Fourier basis $e_k$, we may qualitatively discuss three distinct cases. First, consider the case where $\left|R e_k \right|^2 \sim \mathcal{O}(1)$. Typical examples are the measurements of individual locations or sub intervals of $I$. In this case $\epsilon_R$ is comparable to $\epsilon$. The second case are measurements that suppress small scales, as for example integration over an interval or convolution with a spatially extended kernel such as a point spread function. In case of integration, we get that $\left|R e_k\right|^2 \propto \nicefrac{1}{|k|^2}$ and thus $\epsilon_R$ becomes smaller then $\epsilon$. In these two cases the discretization error of the observable is comparable or smaller then $\epsilon$ and thus a small discretization error for the field is sufficient for a consistent reconstruction. The third case are measurement operators which amplify small scale structures. One important special example are measurement operations involving spatial derivatives. The action of the derivative on $e_k$ leads to a multiplicative factor proportional to $k$ and therefore $\left|R e_k\right|^2 \propto |k|^2$. Care must be taken in this case since $\epsilon_R$ may become large or even infinite even though $\epsilon$ is small. This also shows that not all combinations of $g$ and $R$ lead to an inference problem that allows for a consistent finite dimensional representation.

Nevertheless, given a consistent combination of $g$ and $R$, there always exists a cutoff $K$, that can be chosen prior to the reconstruction, for which the error between the discrete representation and the continuous inference problem becomes small.

\subsubsection{Unknown prior spectrum}
In this work, however, we aim to infer the prior correlation structure, specifically the form of $g$, in addition to $\xi$, from the observed data. This poses a problem, since we cannot deduce a sensible choice for $K$ a priori. In some cases, the measurement setup allows to provide an estimate for the small scale / asymptotic behaviour of the true spectrum from the observed data, and therefore allows to set $K$ accordingly. In many cases, however, such an estimate is not feasible without performing the full reconstruction. One approach to resolve this issue is that after a reconstruction with a chosen $K$, a new reconstruction using a larger cutoff $K' > K$ is performed, and both results are compared. If the two results obtained this way are similar, specifically if the difference between the large and small reconstruction for $s$ as well as the observable $R s$ are small, we may conclude that the chosen discretization sufficiently resolves the true underlying process. If this is not the case, the reconstruction has to be re-run with an even larger cutoff and the procedure has to be repeated, until the deviations become small enough. This ensures that the reconstruction is consistent with the infinite dimensional problem, assuming that the true observed process has a further decaying spectrum for modes above the cutoff.

\subsection{Higher dimensional representation in space-time}\label{sec:spacetimedom}
In order perform inference in a space-time setting we define a $d+1$ dimensional space $\Omega$, where $d$ is the number of spatial dimensions. Specifically we define
\begin{equation}
	\Omega \equiv \left[0, T\right] \otimes \left(\bigotimes_{i=1}^d \left[0, B_i\right] \right) \ ,
\end{equation}
and impose periodic boundary conditions along each axis of $\Omega$. This allows for a multi-dimensional Fourier series expansion of random processes $s \in L^2(\Omega)$, that are statistically homogeneous in space and time, in direct analogy to the one dimensional setting. Furthermore we may label coordinates on $\Omega$ via $x = \left(t, \mathbf{x}\right)$, and additionally $k = \left(\omega, \mathbf{k}\right)$ labels the associated Fourier modes.

The periodic boundary conditions introduce possibly unwanted correlations between the boundaries, in particular along the time axis. To avoid this, we extend the time domain to $2 T$, to be twice the size of the observed domain, and ensure by construction of the prior of $g$ that the process becomes uncorrelated for moments in time with a distance greater than $T$. Similar to the procedure of finding an appropriate discretization, we may ensure that the posterior of $g$ has sufficiently decreased within the interval $[0,T]$ after the reconstruction. If this requirement is not met, the time domain has to be enlarged even further and the reconstruction is performed again. For the sake of simplicity, in all examples of this work, we keep the periodic boundaries in the spatial dimensions. However, these may be omitted as well by an analogous extension of the space.

\section{Prior}\label{sec:prior}
In the introduction we proposed three different concepts we aim to encode into our prior model, particularly into the prior of the correlation structure, defined via $g$.
These are
\begin{itemize}
	\item Statistical space-time homogeneity
	\item (Spatio-) temporal causality
	\item Locality.
\end{itemize}
The first concept is already satisfied by construction as we assume that the correlation structure is diagonal in the Fourier representation and thus fully specified via the Fourier spectrum $g$. We may associate a Linear operator $G$ with $g$, by Fourier transforming the diagonal Matrix $\widehat{G}$ defined in Eq.\ \eqref{eq:fourier_basis} as
\begin{equation}
	G(x,x') = G(x-x') \equiv \left(\mathcal{F} \widehat{G} \mathcal{F}^\dagger\right)(x,x') = \sum_{\omega = - \infty}^\infty \sum_{\mathbf{x} = - \infty}^\infty g^{\omega \mathbf{k}} \ e^{2 \pi i \left(\omega \frac{t-t'}{2 T} + \mathbf{k} \cdot \frac{\mathbf{x} - \mathbf{x}'}{\mathbf{B}}\right)} \ ,
\end{equation}
where $\mathbf{B} = \left(B_1, ..., B_d\right)$ denotes the length of the spatial domain along each axis and $\cdot$ denotes the scalar product. This is the representation of $G$ in space-time coordinates. Indeed we find that
\begin{equation}
	s^x = \left(\mathcal{F} \widehat{G} \xi\right)^x = \int_{\Omega} G(x-x') \ \xi^{x'} \ \mathrm{d}x' \quad \text{with} \quad \xi^{x'} \equiv \left(\mathcal{F} \xi \right)^{x'} \ .
\end{equation}
This equivalence allows for a convenient physical interpretation of $G$ and $\xi$. Assume that $s$ models the deviations of a physical system from its steady state, induced via an external force, an excitation. Furthermore assume that the response of the system to such an excitation is stationary in space and time, i.E. is the same irrespective of where the excitation happened. This implies that $\xi$ plays the role of the external excitation, while $G$ models the stationary response (also called Green's function) of the system to $\xi$. This interpretation is purely artificial at this point, however it motivates the prior concepts that we aim to include into $G$ as they are fundamental for the response of a physical system.

The second concept, causality, is introduced via an additional constraint on $G$. As $G$ should model the response of the system to the excitations $\xi$, it should not contain a response at times before an excitation happens. This can be formulated mathematically as
\begin{equation}\label{eq:caustgreens}
G_{\mathrm{c}}(x-x') = \Theta(t - t') \ G(x - x') \ ,
\end{equation}
where $\Theta$ denotes a step function in time. This trivially ensures that no response happens before an excitation. Technically, as we imposed periodic boundary conditions in time, the constraint has to be modified such that
\begin{equation}\label{eq:caustgreensperiodic}
G_{\mathrm{c}}(x-x') = \Theta(t - t') \ \Theta(T- (t - t')) \ G(x - x') \ .
\end{equation}
The additional step function ensures that an excitation at $t'$ can only cause a response within the interval $[t',T+t']$ but leaves later times unaffected. Together with the expansion of the space by a factor of $2$ as discussed in section \ref{sec:spacetimedom} it ensures that excitations at $T$ do not wrap around the space to cause a response at $t=0$.

\begin{figure*}[htp]
	\centering
	\includegraphics[scale=0.6, angle=0]{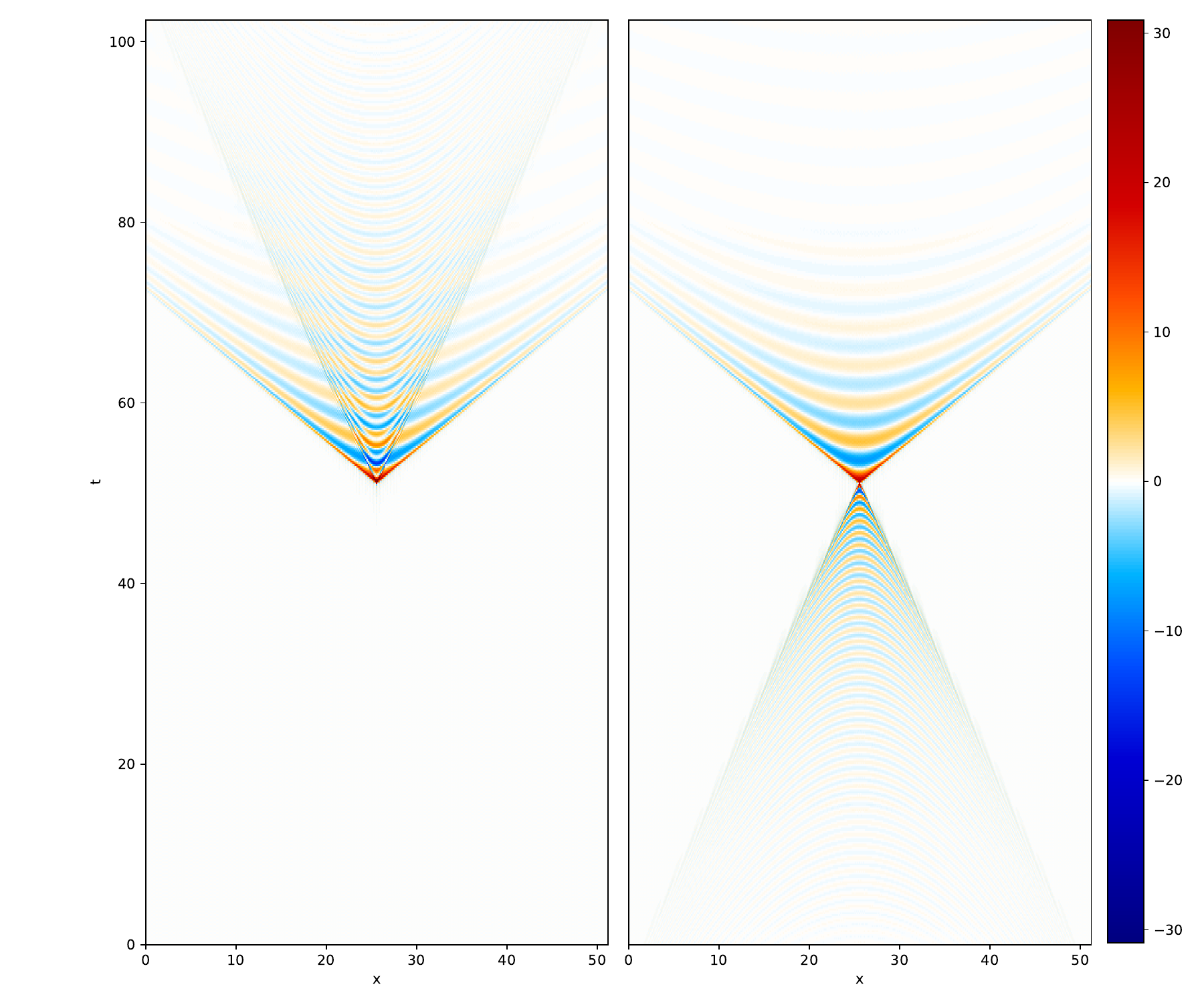}
	\centering
	\caption{{\bf Left:} Causal response of a system defined as a combination of two damped harmonic oscillators with different masses. {\bf Right:} Anti-causal response of the system where one of the oscillations travels backwards in time.}\label{fig:anticausal}
\end{figure*}

In space-time, the physical concept of causality also implies a maximal finite propagation speed of interactions. This leads to propagation within a light cone, as depicted in figure \ref{fig:anticausal}. Therefore, given a maximal propagation speed $c$, we can restrict the reconstruction to Greens functions that are non-zero only within the light cone. We can implement this constraint by extending eq.\ \ref{eq:caustgreens} as
\begin{equation}
G_{\mathrm{lc}}(x-x') = \Theta(l^2) \ G_{\mathrm{c}}(x - x') \ ,
\end{equation}
where
\begin{equation}
l^2 = (t - t')^2 - (\mathbf{x}-\mathbf{x}')^\dagger \mathbf{C}^{-1} (\mathbf{x}-\mathbf{x}') \ .
\end{equation}
This ensures that the propagator is only non-zero within the light cone.
In general we might expect different maximal propagation speeds in different directions, and consequently $\mathbf{C}$ becomes a symmetric tensor. In case propagation is isotropic in space with speed $c$ we have $\mathbf{C}_{ij} = c^2 \delta_{ij}$.
Note that even in cases where we do not know $\mathbf{C}$, for example if the medium in which the interaction is realized is unknown, such a constraint can also be useful if we elevate $\mathbf{C}$ to be an additional unknown parameter that has to be inferred. Since $\mathbf{C}$ is assumed to be the same for all scales, an inference algorithm can use large scale information, where the signal to noise ration (SNR) is usually higher, to determine $\mathbf{C}$, which then effectively increases the SNR for smaller scales. A detailed description how $\mathbf{C}$ enters the reconstruction can be found in appendix \ref{ap:cone}.

\begin{figure*}[htp]
	\centering
	\includegraphics[scale=0.7, angle=0]{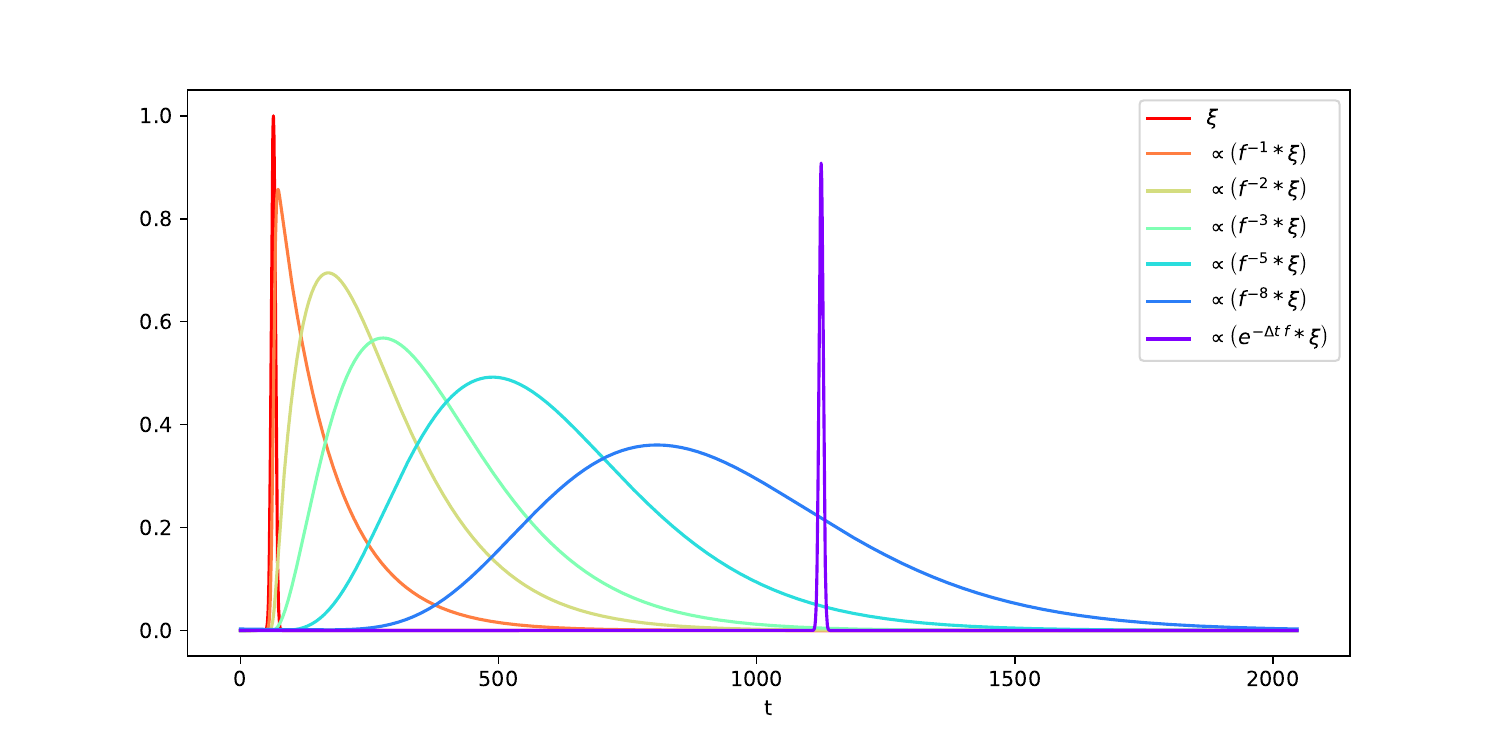}
	\centering
	\caption{Excitation $\xi$ and responses of various systems defined in terms of powers of $f$, with $f_t = \partial_t + \gamma$ and $\gamma$ small. We see that higher order derivatives result in apparent non-local responses. Since time derivatives are the generators of temporal translation, the case where the response is the exponential of $- \Delta t \ f$ leads to translations by $\Delta t$.}\label{fig:order}
\end{figure*}

In order to encode the last concept, locality, we first have to revisit some properties of Green's functions. Consider a system, undisturbed by external excitation, that can be described via
\begin{equation}\label{eq:model}
\mathcal{L} s = 0 \ ,
\end{equation}
where $\mathcal{L}$ is a linear differential operator. The response $G$ of such a system to an external force has to fulfill
\begin{equation}
\left( \mathcal{L} G \right)(x-x') = \delta(x-x') \ .
\end{equation}
In order to give rise to a homogeneous $G$, $\mathcal{L}$ also has to be homogeneous and therefore is diagonal in Fourier space
\begin{equation}\label{eq:linophom}
\mathcal{L}(x-x') = \left(\mathcal{F} \widehat{f} \mathcal{F}^\dagger\right) \ ,
\end{equation}
where we also introduced the diagonal Fourier space representation of $\mathcal{L}$ denoted via the complex Fourier coefficients $f^k$. Consequently $G$ takes the form
\begin{equation}
G^k_{k'} = \frac{1}{f^k} \delta^k_{k'} \ ,
\end{equation}
and therefore the eigen-spectrum of $G$ reads
\begin{equation}\label{eq:spec}
g^k = \frac{1}{f^k} \ .
\end{equation}
It turns out that locality can be encoded more intuitively in terms of a prior for $f$. As we can see in Figure \ref{fig:order}, the locality of a response is related to the order in the derivatives of the differential operator $\mathcal{L}$. Low order derivatives result in an almost instantaneous response of the system while higher order derivatives lead to an apparent non-local response. Therefore we seek to formulate a prior for $f$ such that lower order derivatives are a priori favoured. Nevertheless it should be possible that $f$ can deviate from this assumption if there is enough evidence in the data to support this. We impose a certain degree of smoothness for $f$, i.E.\ we want that two modes $f^k$, $f^{k'}$ are correlated, where the correlation decays as $|k-k'|$ increases. To this end we define the complex modes $f$ as
\begin{equation}
	f^k = f_r(k) + i f_i(k) \quad \text{with}  \quad k \in \mathds{Z} \ , \quad f_r, f_i \in L^2(\mathds{R}) \ .
\end{equation}
In words, the real and imaginary part of the modes $f^k$ are defined via two square integrable functions $f_{\nicefrac{r}{i}}$ which get evaluated at the integer spaced Fourier locations $k$. We place a Gaussian process prior on both functions $f_r$ and $f_i$ of the form
\begin{equation}\label{eq:gausssmoothn}
P(f_{\nicefrac{r}{i}}) = \mathcal{G}(f_{\nicefrac{r}{i}},T) = \frac{1}{\left| 2 \pi T \right|^{\frac{1}{2}}} e^{- \frac{1}{2} f_{\nicefrac{r}{i}}^\dagger T^{-1} f_{\nicefrac{r}{i}}} \ ,
\end{equation}
where the exponential factor reads
\begin{equation}\label{eq:smoothn}
- \frac{1}{2} f_{\nicefrac{r}{i}}^\dagger T^{-1} f_{\nicefrac{r}{i}} = -\frac{1}{2 \sigma^2} \int \left|\left(\gamma - \triangle_k\right)f_{\nicefrac{r}{i}}(k)\right|^2 dk \quad \gamma, \sigma > 0 \ ,
\end{equation}
where $\Delta_k$ denotes the Laplace operator, $\sigma$ is an overall scaling parameter that steers the strength of this prior, and $\gamma$ is a low frequency cutoff to ensure that the prior is proper. For further details see \cite{2013PhRvE..87c2136O,2017PhRvE..96e2104F}.
The covariance function associated with $T$ takes the form
\begin{equation}
	T(k,k')  = \frac{\sigma^2}{2} \sqrt{\frac{\pi}{2 \gamma^3}} \ \left(1 + \sqrt{\gamma} \ \left|k - k'\right|\right) \ e^{- \sqrt{\gamma} \ \left|k - k'\right|} \ .
\end{equation}
Therefore the real and imaginary part of the Fourier modes $f^k$, which are just the functions $f_{\nicefrac{r}{i}}$ evaluated at $k \in \mathds{Z}$, are defined to be two independent, infinite dimensional Gaussian random vectors with zero mean and covariance
\begin{equation}
	T_{k k'} = T(k,k') \ , \quad k, k' \in \mathds{Z} \ .
\end{equation}

All in all, we can combine the above concepts to end up with a generative description of our prior which takes the form
\begin{equation}\label{eq:generativemod}
s(a) = G_{\mathrm{lc}} \ \xi = \mathcal{F} \ \widehat{g_{\mathrm{lc}}} \ \mathcal{F}^\dagger \ \xi \ ,
\end{equation}
where $\widehat{g_{\mathrm{lc}}}$ denotes a diagonal matrix in Fourier space with $(g_{\mathrm{lc}})^k$ on its diagonal and $a = \left(f,\xi, \mathbf{C}\right)$ denote the quantities of interest in the generative process. Furthermore
\begin{equation}\label{eq:eigengreenf}
g_{\mathrm{lc}} = \mathcal{F}^\dagger \ \widehat{X} \ \mathcal{F} \frac{1}{f} \quad \text{with} \quad X^x = \Theta(T-t) \Theta(t) \ \Theta(l^2) \ ,
\end{equation}
with $f$ being a priori distributed according to eq.\ \eqref{eq:gausssmoothn}.

\subsection{Comparison to Mat\`{e}rn type and other parametric kernels}
There exists a vast literature about Gaussian process priors with a stationary covariance \cite{genton2001classes, rasmussen2003gaussian} which discuss a great variety of different forms of covariance functions. Two important classes are squared exponential, and the Mat\`{e}rn class of covariance Functions. As a stationary covariance allows for a diagonal representation in Fourier space, it makes sense to compare the spectra of the associated operators. Squared exponential kernels imply that the spectrum takes the form
\begin{equation}
	g^k \propto e^{- \gamma |k|^2} \quad \gamma > 0 \ .
\end{equation}
While such kernels are very popular as they are particularly easy to implement and use in practice, the quadratic-exponential suppression of small scale structures often appears to be non-physical. A physically better motivated type of spectrum is provided by Mat\`{e}rn covariances which give rise to spectra of the form
\begin{equation}
	g^k \propto \frac{\alpha}{\left(\beta + |k|^2\right)^\gamma} \quad \alpha, \beta, \gamma > 0 \ .
\end{equation}
As many physical processes can be well approximated as a power-law, this parametrization provides a more sensible statistical structure. In addition the large-scale cutoff ensures that the process is well defined as the variance remains finite for all $k$. We notice that the denominator of this spectrum is very well represented by the prior process we impose on $f$ and thus the Mat\`{e}rn class of covariances is represented in our prior assumptions. In contrast to a fixed Mat\`{e}rn covariance function, however, the form of $f$ remains unknown prior to the reconstruction and thus is inferred to match the observed data. In addition, a non-parametric process for $f$ appears to be more flexible in modeling deviations from a (possibly idealized) power-law shape of the spectrum.

\subsection{Prior distributions for excitations}
So far we restricted the discussion to independent Gaussian distributed excitations $\xi$ which, for a given fixed $G$, give rise to a Gaussian process for $s$.

From the physical perspective of $s$ being the result of a dynamic response $G$ to external excitations $\xi$, it is not necessary that $\xi$ is Gaussian distributed. In fact, in many applications it might be more realistic to define a different prior for the excitations. In order to demonstrate the implications of a non-Gaussian prior on $\xi$, in section \ref{sec:application}, we show results for the inference problem in cases where $\xi$ is distributed according to an inverse-gamma distribution at each location in space-time. This prior is typically used as a sparsity prior, and in our case results in a system that is subject to sparse external excitations. Of course, many other prior distributions are also reasonable and important, but for the sake of simplicity we stick to $\xi$ being either Gaussian or inverse-gamma distributed in the examples. Note that even though physically motivated, exchanging the prior of $\xi$ to be an inverse-gamma distribution is non-trivial in the continuum limit. The goal of the examples is to demonstrate the applicability of this method also for non-Gaussian excitations, and therefore a rigorous mathematical treatment is beyond the scope of this work. In the case of inverse Gamma excitations, we therefore revert to the discrete representation of the process and leave the continuous treatment to future research.

\section{Inference}\label{sec:inference}
In section \ref{sec:wf} we already discussed the inference problem in the case of a linear measurement and a Gaussian prior with known covariance. Now, with the appropriate prior for the covariance at hand, we can set up the task of inferring the covariance together with the field, given observational data about the field. To this end, consider again a linear measurement as defined in eq.\ \eqref{eq:data_model} which in terms of $a$ takes the form
\begin{equation}\label{eq:datamodel}
d = R \ s(a) + n \ ,
\end{equation}
where $s(a)$ is given via Eq.\ \eqref{eq:generativemod}. If we assume $n$ being Gaussian distributed with zero mean and covariance $N$ we get that the joint distribution of $(d,a)$ reads
\begin{equation}\label{eq:joint}
P(d,a) = \mathcal{G}(d-R \ s(a),N) \  \mathcal{G}(f,T) \ P(\xi) \ .
\end{equation}
The posterior $P(a|d)$ is proportional to the joint distribution up to a factor that only depends on $d$ since
\begin{equation}\label{eq:post}
P(a,d) = P(a|d) \ P(d) \propto P(a|d) \ .
\end{equation}
This posterior is intractable, due to the non-linear dependency of $s$ on $a$. Consequently the corresponding inference problem cannot be solved analytically and we have to rely on a numerical approximation. 

There exist a variety of different approximation techniques for Bayesian inverse problems ranging from point estimations such as the maximum a posterior (MAP) estimate, over variational approximations, to posterior sampling techniques such as Markov-Chain Monte Carlo (MCMC) \cite{brooks2011handbook, cotter2013mcmc} and Hybrid Monte Carlo (HMC) \cite{duane1987hybrid}. As shown in \cite{2019arXiv190111033K}, MAP tend to perform poorly in the task of reconstructing the excitations together with the prior correlation structure, as uncertainty information is vital to correctly estimate the prior statistics, which are missing in point estimates. While MCMC and HMC algorithms are very attractive due to their theoretical guarantees to converge to the true posterior statistics, they tend to become expensive for many astrophysical field inference applications compared to simpler, less expressive approaches. Therefore, in this work, we use a variational approximation algorithm called Metric Gaussian Variational Inference (MGVI) \cite{2019arXiv190111033K} where we approximate the true posterior with a Gaussian distribution in order to get an estimate for the mean and the covariance of the posterior. As shown in \cite{2019arXiv190111033K}, MGVI provides an accurate estimate of the first moment (i.E. the posterior mean) as well as a tight lower bound on the second moment (posterior covariance) when compared against HMC techniques, while being substantially faster.

\subsection{Variational Inference}
In general, variational inference can be described as the task of approximating one probability distribution $P(x)$ for some quantity $x$ with another distribution $Q_\sigma(x)$ which, in addition, is defined up to a set of parameters $\sigma$. Approximation is then achieved via minimizing the Forward Kullbach-Leibler divergence (KL) with respect to $\sigma$. The KL is defined as
\begin{align}
\mathrm{KL}(Q_\sigma,P) &\equiv \int Q_\sigma(x) \log\left(\frac{Q_\sigma(x)}{P(x)}\right) \mathrm{d}x \notag\\ &=
\left< H_P \right>_{Q_\sigma} - \left< H_{Q_\sigma} \right>_{Q_\sigma}\ ,
\end{align}
where we also introduced the so called Information Hamiltonian $H$ defined as
\begin{equation}
H_P = - \log(P) \ .
\end{equation}

In our case, the approximate distribution $Q$ is chosen to be a Gaussian distribution in $a$ with mean $m$ and covariance $A$. For many relevant inference problems, and also for the one studied in this work, this approximation cannot be performed analytically as typically $\left< H_P \right>_{Q_\sigma}$ cannot be calculated analytically. Therefore, as discussed in section \ref{sec:discretization}, we choose an appropriate discretization for the space-time domain $\Omega$, and perform a variational approximation of the corresponding discrete problem. It turns out that in order to achieve a reasonable resolution in space-time, the inference problems can become very high dimensional. As an example, in section \ref{sec:application}, we show an application for a discretized 1+1 dimensional space-time with a resolution of $256 \times 200$ pixels. This yields that the number of dofs in $a$ is $\approx 2 \times 256 \times 200 \propto 10^5$ and consequently the number of entries in $A$ are $\propto 10^{10}$ which renders an explicit representation of $A$ on a computer to be inefficient generally.
Therefore, following \cite{2019arXiv190111033K}, we avoid an explicit representation by setting it to be equal to the inverse Fisher Metric $\mathcal{M}^{-1}$ \cite{amari2007methods} of the posterior, evaluated at $m$. For the posterior distribution as defined in eqs.\ \eqref{eq:joint} and \eqref{eq:post}, together with a Gaussian prior for $\xi$ with zero mean and unit covariance, we get that $\mathcal{M}$ takes the form
\begin{align}\label{eq:fisher}
\mathcal{M} = \left[\left(\frac{\partial s}{\partial a}\right)^\dagger R^\dagger N^{-1} R \frac{\partial s}{\partial a} +  \left(\begin{matrix}
\mathds{1} & 0  \\ 0 & T^{-1}
\end{matrix}\right)\right]_{a = m} \equiv \left(\tilde{R}_m\right)^\dagger N^{-1} \tilde{R}_m + S^{-1} \ .
\end{align}
Using the definition of $s$ (eq.\ \eqref{eq:generativemod}) we get that
\begin{align}
\frac{\partial s}{\partial \xi} &= \mathcal{F}^{-1} \ \widehat{g} \ \mathcal{F} \ , \\
\frac{\partial s}{\partial f } &= \mathcal{F}^{-1} \widehat{\mathcal{F} \xi} \mathcal{F} \ \widehat{X}  \ \mathcal{F}^{-1} \widehat{\nicefrac{-1}{f^2}} \ .
\end{align}
Therefore, together with the definition of $T$ (eq.\ \eqref{eq:smoothn}) we see that $\mathcal{M}$ has an implicit representation, i.e. it can be applied solely using Fourier transformations and diagonal operations, avoiding the explicit storage of this matrix at any point. Consequently the application of $A = \mathcal{M}^{-1}$ is achieved via linear solvers such as the conjugate gradient method.
In addition, the structure of $\mathcal{M}$ allows for an efficient sampling of the approximate distribution $Q$. Specifically we may draw a random realization of $Q$ as
\begin{equation}
	a^* = m + \mathcal{M}^{-1} \left(\left(\tilde{R}_m\right)^\dagger n^* + b^*\right) \ ,
\end{equation}
where $n^*$ and $b^*$ are independent samples drawn from the noise statistics and the joint prior distribution with covariance $S$, respectively. 

This is another important property since the KL involves expectation values w.r.t $Q$, which can be approximated via samples from $Q$. Ultimately, the Fisher metric is also a measure for the local curvature of the KL and therefore enables us to use second order optimization schemes to solve the corresponding optimization problem in $m$. 

As discussed in the previous section, in some cases space-time causality can only be imposed if we also infer the propagation speed $\mathbf{C}$. To do so, we notice that given a prior for $\mathbf{C}$, the above still applies with the extension that $s$ is now also a function of $\mathbf{C}$. Therefore the Fisher metric gets an additional entry with the same structure as in eq.\ \eqref{eq:fisher}, for the corresponding gradient $\nicefrac{\partial s}{\partial \mathbf{C}}$ (See appendix \ref{ap:cone} for further details).

All in all, the solution strategy as defined in MGVI, starting from a random initialized $m$, can be summarized as:
\begin{itemize}\label{alg:mgvi}
	\item Using $\mathcal{M}$, evaluated at the current approximate mean $m$, draw a set of samples $\{a^*\}$ from the approximate distribution $Q$.
	\item With these samples, calculate an estimation for the current value of the KL and its gradient.
	\item Together with the metric $\mathcal{M}$ perform a second order Newton minimization in order to get a new estimate for $m$.
	\item Repeat this procedure by re-evaluating everything using the updated $m$, until convergence.
\end{itemize}

\section{Application}\label{sec:application}
To demonstrate the applicability of our method we apply it to several synthetic data examples. First, we demonstrate the performance of the algorithm in a one dimensional setting, where we only aim to infer the Green's function $G$ of the system, given the excitations and data. In the second example we perform a reconstruction of excitations that are distributed according to a unit Gaussian, as well as the Green's function from data alone, in a 1+1 dimensional setting. In the last example we add another level of complexity via non-Gaussian statistics in the excitations. Specifically we assume the excitations to follow an inverse-gamma distribution at each location in discretized space-time. Hereby we perform source detection, the task of inferring sparse excitations at various locations in space-time, in a case where also the Green's function is unknown. Throughout all examples, the prior model for $G$ follows the one described in section \ref{sec:prior}. Specifically, we describe $G$ in terms of $f$ and the propagation speed $\mathbf{C}$. We place a flat prior on the logarithm of $\mathbf{C}$ to ensure positivity of the propagation speed.

\subsection{Implementation details}
All examples were implemented using the NIFTy software package in its version 5 \cite{2019ascl.soft03008A}. We included an implementation of the here introduced prior structure for the Green's function into this publicly available package.

Throughout all examples, to solve the optimization problem associated with the MGVI algorithm, we realize the algorithm described in \ref{alg:mgvi} with 30 total steps of the entire loop, where we draw 10 samples from the approximation $Q$ to estimate the KL during optimization. For posterior analysis, we use 50 approximate posterior samples.

For the first, one dimensional example, the optimization can be performed within less then a minute, while for the other two examples the total runtime is around 10 minutes on a standard laptop. The runtime and scaling of MGVI with model complexity as well as dimensionality is described in great detail in \cite{2019arXiv190111033K}.

\subsection{Temporal evolution}
\begin{figure*}[htp]
	\centering
	\includegraphics[scale=1., angle=0]{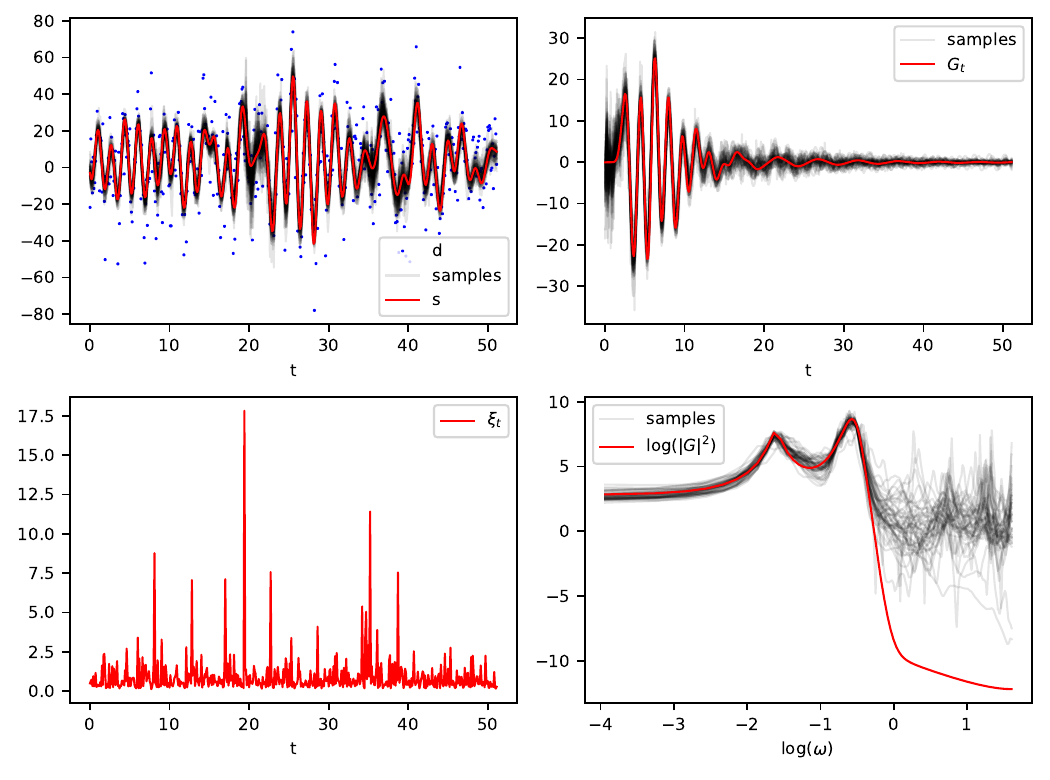}
	\centering
	\caption{{\bf Top:} On the left we depict the signal (red line), the data (blue dots) as well as 50 over-plotted posterior samples (gray). The right panel shows the synthetic propagator (Greens function) in the temporal domain (red) as well as corresponding posterior samples. {\bf Bottom:} Left: Excitation field used to generate the signal. Right: Natural logarithmic spectrum of the synthetic propagator (red) and corresponding posterior samples.} \label{fig:1drec} 
\end{figure*}

In our first application, shown in figure \ref{fig:1drec}, we aim to demonstrate the inference of the dynamics encoding field $f$ alone in a one dimensional setting where there is only a temporal evolution to be reconstructed. The excitation field $\xi$ is known during inference. The hyper-parameters of the prior for $f$ (see Eq.\ \eqref{eq:gausssmoothn}) are set to $(\sigma, \gamma) = (2.9, 1.7)$. We generate synthetic data according to eq.\ \ref{eq:datamodel}, with $R$ being the identity. The excitations were drawn from an inverse gamma distribution to model known, sparse excitations of the system (e.g.\ in a laboratory setting where the unknown system is driven via sparse excitations). The synthetic signal $s$ as well as corresponding data $d$ is shown in the top-left panel of figure \ref{fig:1drec}. The dynamic operator used to generate signal and data is of the form:
\begin{equation}\label{eq:1ddiffop}
\mathcal{L} = (\partial_t^2 + m^2 + \gamma \partial_t)^5 \ (\partial_t^2 + \tilde{m}^2 + \tilde{\gamma} \partial_t) \ ,
\end{equation}
with $(m,\gamma,\tilde{m},\tilde{\gamma}) = (0.6,0.23,0.2,0.03)$. The results of the reconstruction are shown in figure \ref{fig:1drec}.

We see that the reconstruction of the Green's function is in agreement with the ground truth in the temporal domain, within uncertainties. Due to the fact that the reconstructed dynamics is uncertain, the recovered signal also has uncertainty although the excitations are known. The reconstructed Green's function indicates that it is indeed possible to reconstruct an apparent non-local response of the system (due to higher order derivatives in this setup) since the true propagator as well as its reconstruction show oscillations that grow in the beginning of the propagator before decaying exponentially. We also notice that there is relatively high uncertainty in the first timesteps of the response $G_t$. This is caused by the low initial response to excitations of the true propagator. The initial part of the reconstructed propagator is purely dominated by noise and thus only constrained up to the noise level (the standard deviation of the noise $\sigma_n$ is set to be $\sigma_n = 15$ and the temporal domain is discretized via 512 equidistant pixels).

We also notice that the posterior solution for the spectrum levels out for high-frequency modes (large $\omega$) below the signal to noise ratio, while the true spectrum continues to decay (ultimately also the true spectrum levels out in the numerical example due to the finite size of the considered space). The uncertainty increases in this region, but not enough to capture the true solution. Here we notice the limitations of the variational inference, which provides a local approximation of the posterior with a Gaussian. Consequently the true uncertainty might be underestimated, as in this case. However this deviation occurs orders of magnitudes below the peak of the spectrum and therefore has only a barely visible effect on the reconstruction of the signal. One way to allow for a better extrapolation to higher frequencies would be to provide a more restrictive prior for the dynamics encoding field e.g.\ by defining it on a polynomial basis which is a more suitable basis for this particular setup as the true dynamics is also described in terms of a polynomial. However we aim to provide a general and less restrictive approach here capable of also reconstructing non-polynomial dynamics encoding functions and consequently being less able to extrapolate to regions where we have no information provided by data. 

\begin{figure*}[htp]
	\centering
	\includegraphics[scale=.8, angle=0]{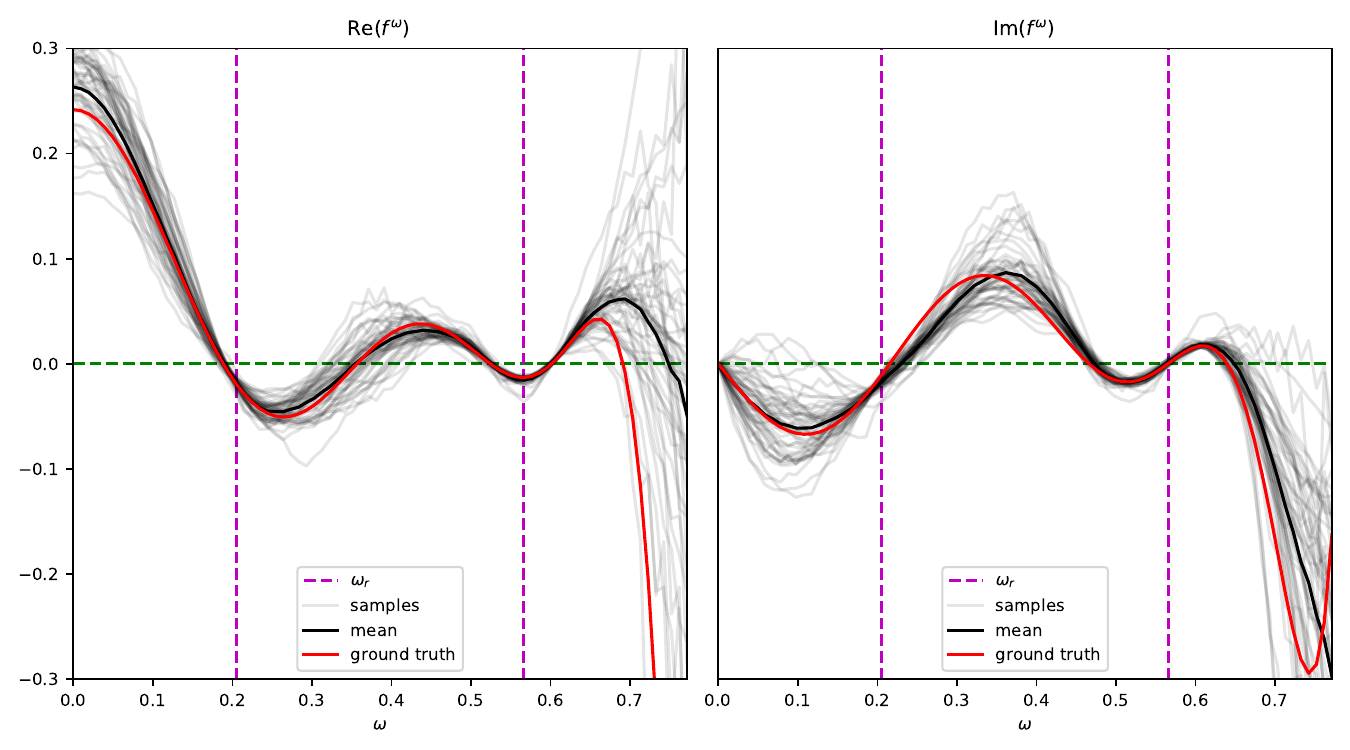}
	\centering
	\caption{Posterior mean (black line) and posterior samples (gray lines) of the real part (left) and imaginary part (right) of the inverse of the propagator spectrum $f^\omega = \nicefrac{1}{g^\omega}$. The red lines indicate the real and imaginary part of the differential operator $\mathcal{L}$ (Eq.\ \eqref{eq:1ddiffop}) used to generate the data of this example. The purple dashed lines indicate the values of the two resonant frequencies $\omega_r$ corresponding to $\mathcal{L}$ where the magnitude of $f$ is smallest and thus the contribution to the observed process $s$ is largest.} \label{fig:1ddiffop} 
\end{figure*}

In addition, in figure \ref{fig:1ddiffop}, we depict the reconstructed real and imaginary part of the inverse of the propagator spectrum and compare it to the spectrum associated to the differential operator $\mathcal{L}$ (Eq.\ \eqref{eq:1ddiffop}) used to generate the mock data. We see that the true spectrum is in agreement with the reconstruction, within uncertainty. Furthermore we notice that the posterior uncertainty is small close to the two resonant frequencies $\omega_r$ (see figure \ref{fig:1ddiffop}) corresponding to the two peaks in the propagator spectrum depicted in figure \ref{fig:1drec}. This is due to the fact that at these frequencies both, the real and imaginary part of the spectrum, are close to zero and thus the magnitude of its inverse is large. Therefore small deviations around these values result in large changes in the corresponding realization of the random process $s$ and therefore the posterior uncertainty has to be small in order to stay consistent with the data. 

\begin{figure*}[htp]
	\centering
	\includegraphics[scale=.8, angle=0]{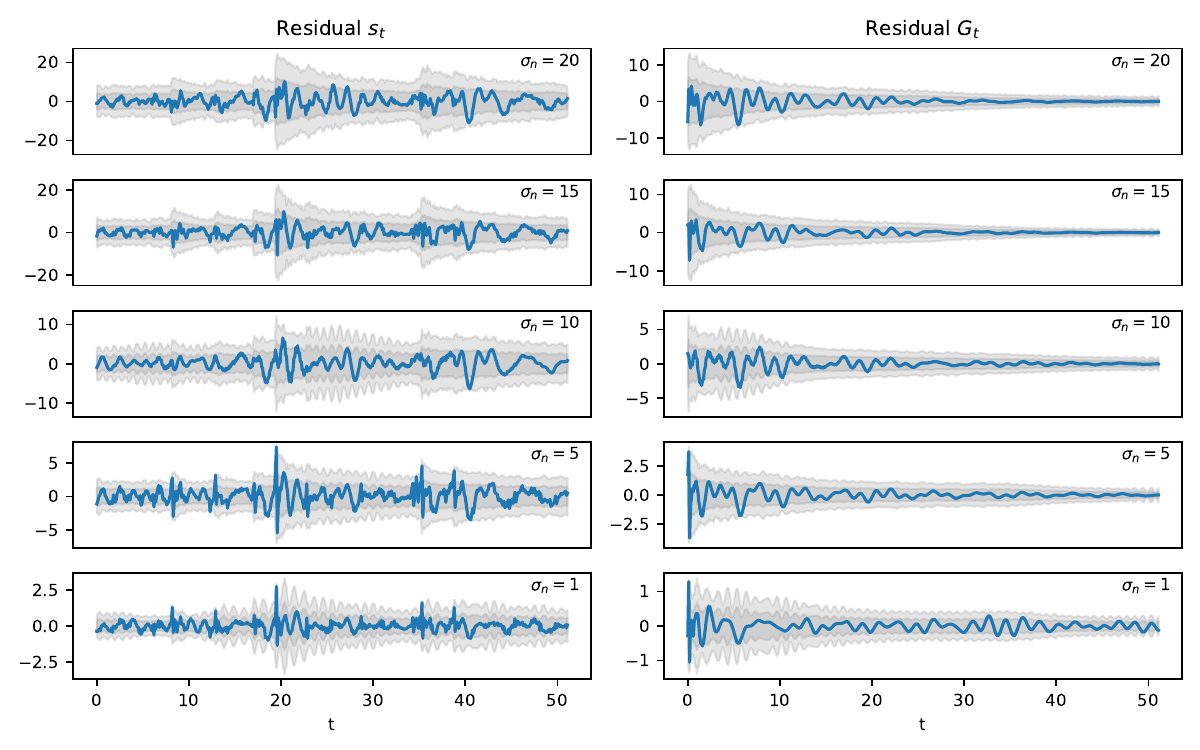}
	\centering
	\caption{{\bf Left:} Residuals between the true signal $s_t$ and the posterior mean of the reconstruction (blue line) as well as the one and two sigma confidence intervals of the corresponding posterior uncertainty. We show those for various different noise standard deviations $\sigma_n$ starting with the highest noise level at the top to the lowest at the bottom. {\bf Right:} Corresponding residuals and confidence intervals for the temporal representation of the dynamic Green's function $G_t$ again for various noise levels. In all inference runs, we seeded the random number generator used for data generation and during reconstruction with the same random seed such that the only difference in these reconstructions is a different $\sigma_n$.} \label{fig:1residual} 
\end{figure*}

To further quantify the reconstruction error, we also investigate the residual as well as the corresponding posterior uncertainty for the signal $s_t$ and the time representation of the Greens function $G_t$ (see figure \ref{fig:1residual}). The residual is defined as the difference between the true solution and the posterior mean of the reconstruction. In addition to the case of $\sigma_n = 15$, we also show the residuals and uncertainties for various other noise levels. For a better comparison, no other changes where made during reconstruction. In particular also the random number generator used to generate the synthetic data as well as the approximate posterior samples during reconstruction was seeded with the same random seed for all runs. We notice that the posterior uncertainty appears to be on a reasonable scale as the residual is within the one or two sigma confidence interval for almost all cases. In addition, we notice that the posterior uncertainty of the signal $s$ is particularly high in regions right after a strong excitation happened. This is due to the fact that the reconstruction of the Green's function $G$, which is the response to these excitations, is most uncertain for the first timesteps. This uncertainty propagates into the uncertainty of $s$.

\subsection{Space-time evolution}\label{sec:con2d}
\begin{figure*}[htp]
	\includegraphics[scale=0.48, angle=0]{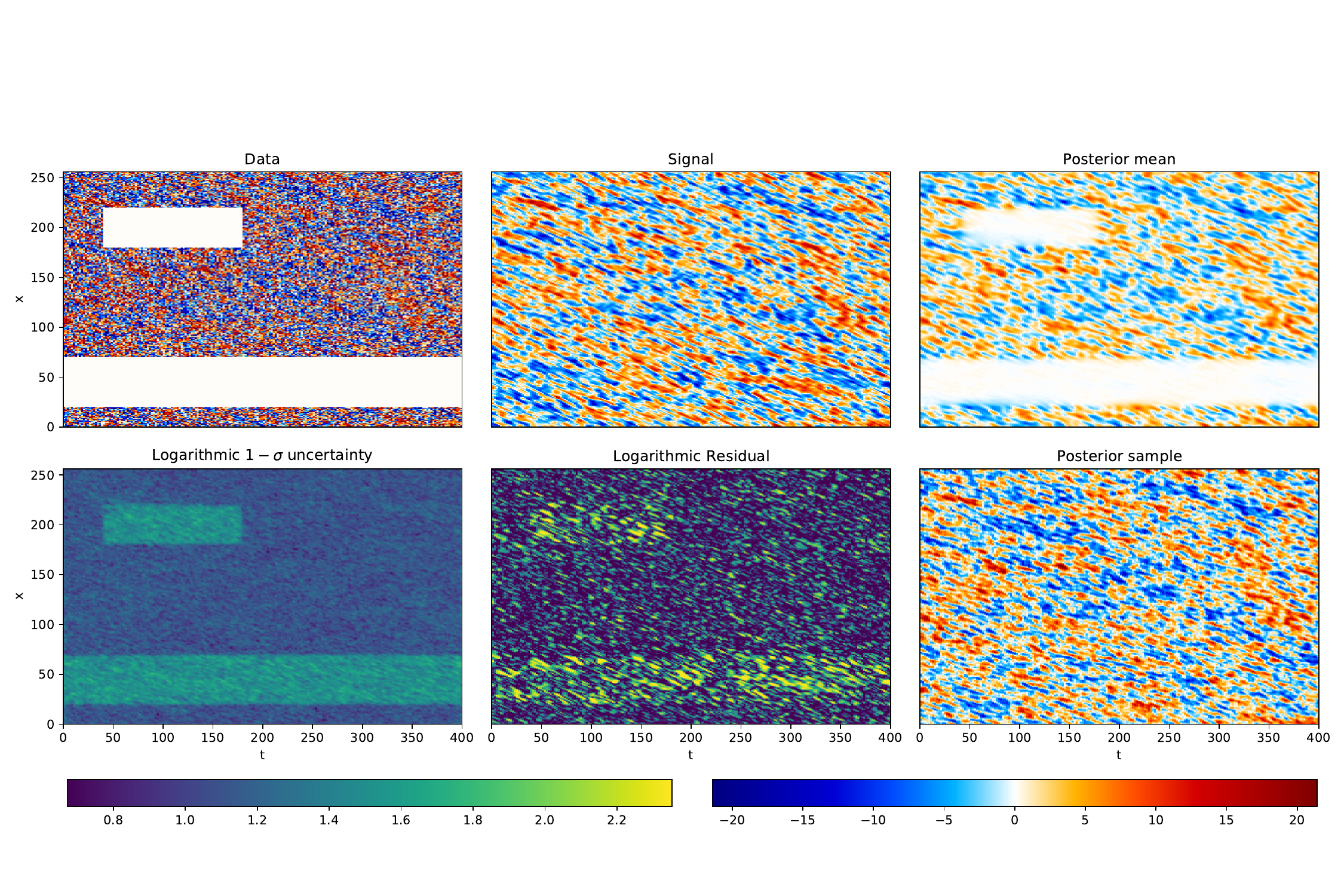}
	\centering
	\caption{{\bf Top:} The left panel shows the spatio-temporal masked and noisy data, drawn from the synthetic signal (middle panel) and the corresponding signal reconstruction (right panel). {\bf Bottom:} Natural logarithmic one-sigma posterior uncertainty (left panel), natural logarithmic residual between the true signal and the posterior mean (middle panel), as well as an approximate posterior sample (right panel).}\label{fig:2dmocksignal}
\end{figure*}

In our second example, shown in figure \ref{fig:2dmocksignal}, we aim to reconstruct the dynamics as well as the excitations in a spatio-temporal (1+1 dimensional) setting from incomplete and noisy observations. We aim to infer the dynamical field $f$, the propagation speed $c$, and the excitations $\xi$ from noisy and incomplete measurements $d$ of the field $s$ alone. The hyper-parameters of the prior for $f$ (see Eq.\ \eqref{eq:gausssmoothn}) are set to $(\sigma, \gamma) = (1.8, 0.5)$. The dynamical system used for the generation of synthetic data is a product of a damped harmonic oscillator and an advection-diffusion generating term. This reads
\begin{equation}
\mathcal{L} = (\partial_t^2 - c^2 \partial_x^2 + m^2 + \alpha \partial_t - \beta \partial_x) \ (\partial_t + \tilde{m}^2 - \gamma \partial_x^2 + \tilde{\beta} \partial_x) \ .
\end{equation}
Furthermore, the excitations are Gaussian distributed with zero mean and unit covariance from which a single realization was drawn and convolved with the synthetic Green's function corresponding to $\mathcal{L}$.
The signal $s$, the data $d$ and the reconstruction of $s$ are shown in figure \ref{fig:2dmocksignal} for a case with $(c,m,\alpha,\beta,\tilde{m},\gamma,\tilde{\beta})=(0.4,0.16,-0.19,-0.05,0.1,0.5,0.2)$. We generate the data according to eq.\ \ref{eq:datamodel} with a linear measurement response, which partially ($\approx 25 \%$) masks the observed region, and Gaussian distributed noise with $\sigma_n = 10$. The space-time is discretized via a regular grid with $256 \times 200$ pixels, respectively.

The reconstruction algorithm is capable of reconstructing the signal in regions where we have observations thereof, while being relatively blind in unobserved regions. Consequently the posterior uncertainty is higher there. In addition, we notice that unlike the posterior mean, posterior samples consistently fill unobserved regions. Although in these regions the samples deviate strongly from the true signal, the information on the statistical properties, inferred from the observed regions, propagates into the unobserved regions due to the assumed statistical homogeneity. Therefore the posterior samples are statistically consistent throughout the entire space-time interval, which is important for posterior analysis. Furthermore we notice that there also exists variance in the statistical properties of the posterior, as can be seen for example in the difference between small scale structures of the posterior mean and the sample displayed in figure \ref{fig:2dmocksignal}. This is due to the fact that also the reconstruction of the statistical properties (described via $f$) is imperfect due to the noisy data and thus subject to uncertainty. This posterior uncertainty about the small scale properties of $f$ results in a variation between different posterior samples of $f$, which ultimately propagates into the statistical properties of the corresponding sample of $s$.

\begin{figure*}[htp]
	\centering
	\includegraphics[scale=0.45, angle=0]{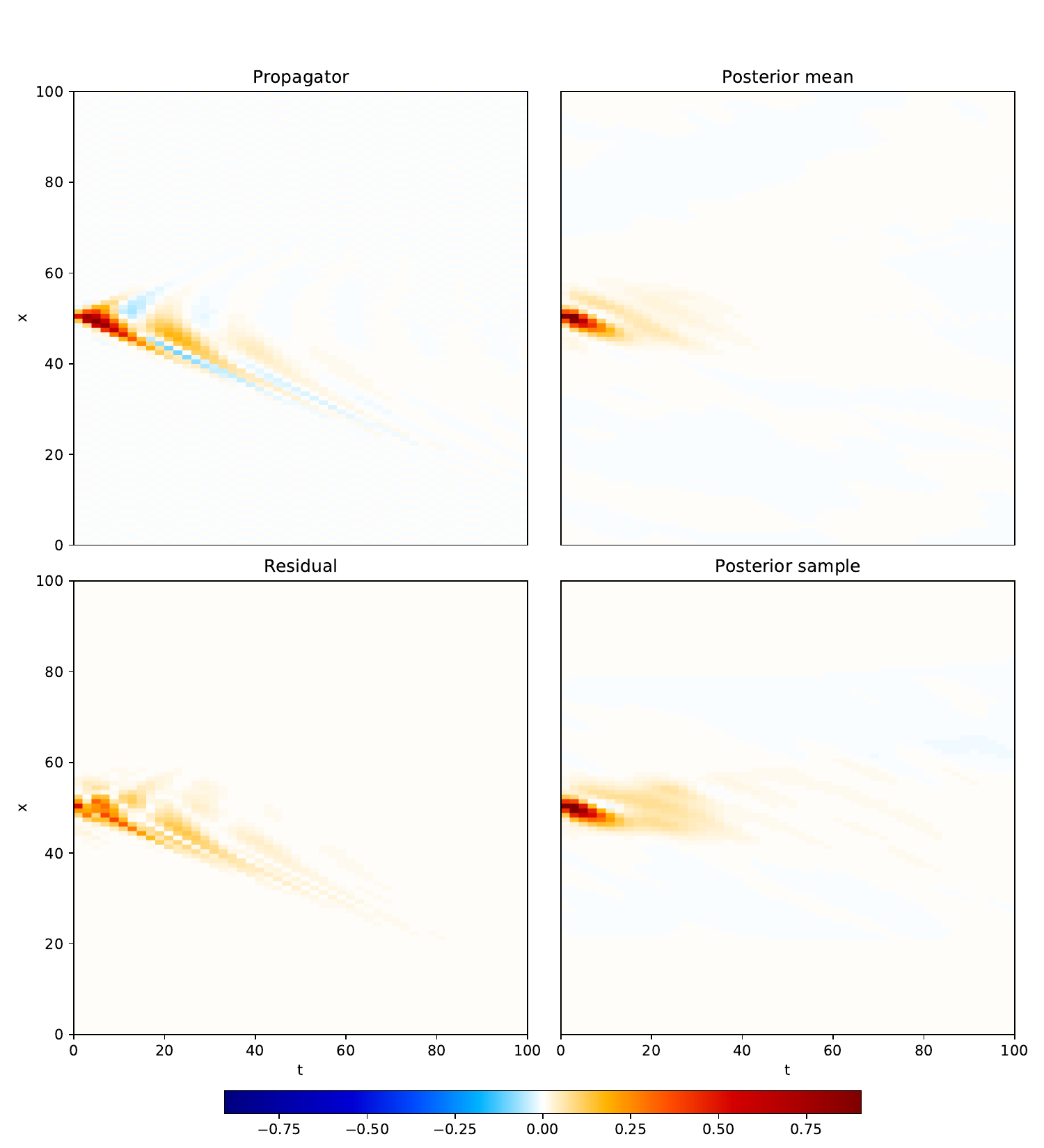}
	\centering
	\caption{{\bf Top:} True Green's function (left) and corresponding posterior mean (right). {\bf Bottom:} Residual of the true Green's function and the reconstruction (left) and a posterior sample for the Green's function (right).}\label{fig:2dmockprop}
	\centering
	\includegraphics[scale=0.45, angle=0]{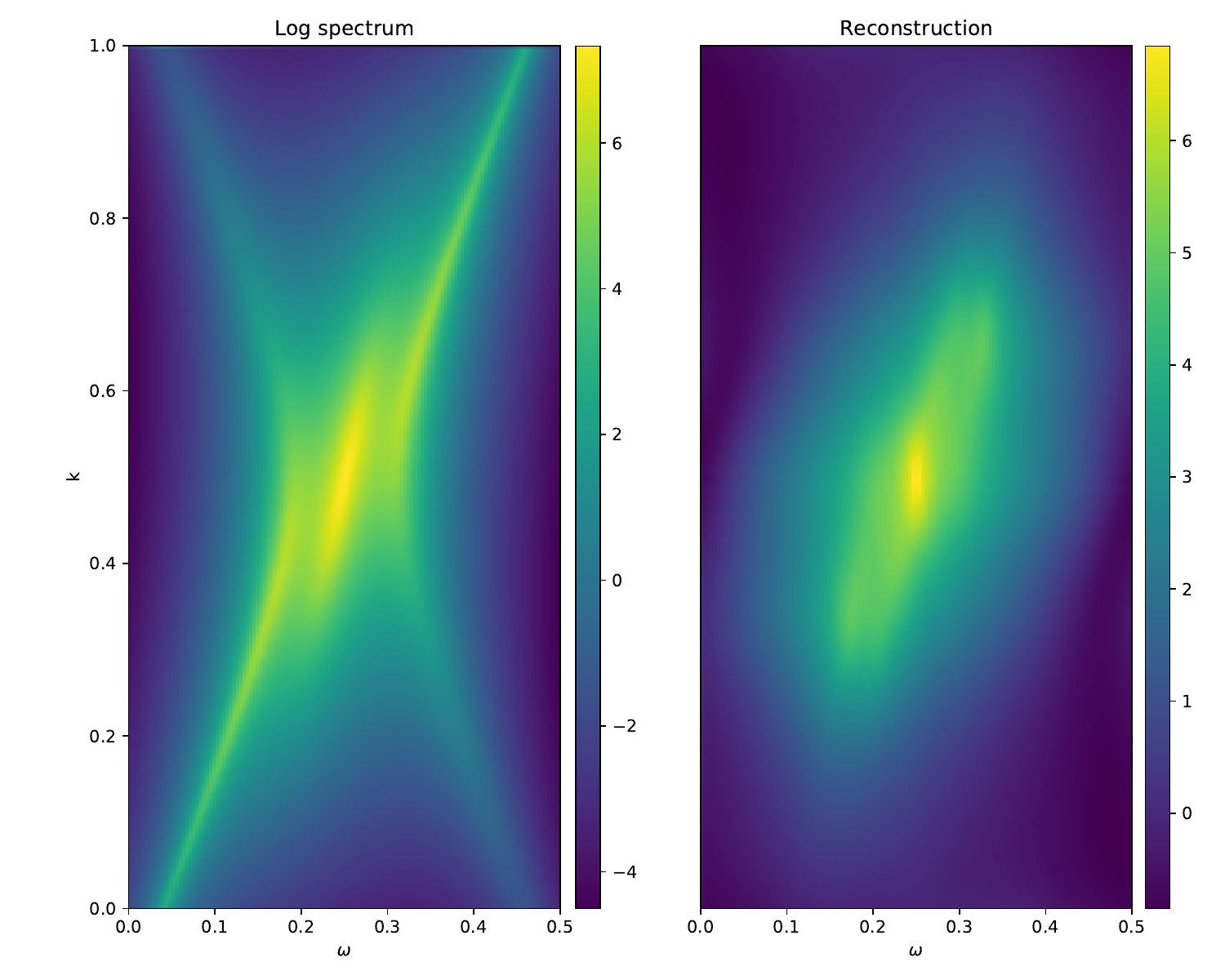}
	\centering
	\caption{Natural logarithmic spectrum of the true Green's function (left) as well as the corresponding posterior mean (right).}\label{fig:2dmockspec}
\end{figure*}

In figures \ref{fig:2dmockprop} and \ref{fig:2dmockspec} we study the posterior properties of the Green's function in more detail. In particular we compare the reconstructed Green's function as well as the corresponding spectrum with the underlying ground truth. The spectrum is comparable to the ground truth in regions with sufficient SNR while it levels out in regions where we have no information given via data. In addition, the reconstructed propagator also shows oscillations consistent with the true propagator. However we notice that modes that propagate ``downwards" are reconstructed well while the weaker ``upwards" propagating modes are not reconstructed due to the fact that they are below the noise level. In addition, we see that deviations in posterior samples of the propagator only occur within a ``cone" and remain zero outside. This is due to the fact that we also reconstruct the maximal propagation speed of the process, which is reconstructed to be $c \approx 0.45 \  (\pm 0.08)$ enclosing the correct value of $c = 0.4$.

\subsection{Source detection}\label{sec:source}
\begin{figure*}[htp]
	\centering
	\includegraphics[scale=0.48, angle=0]{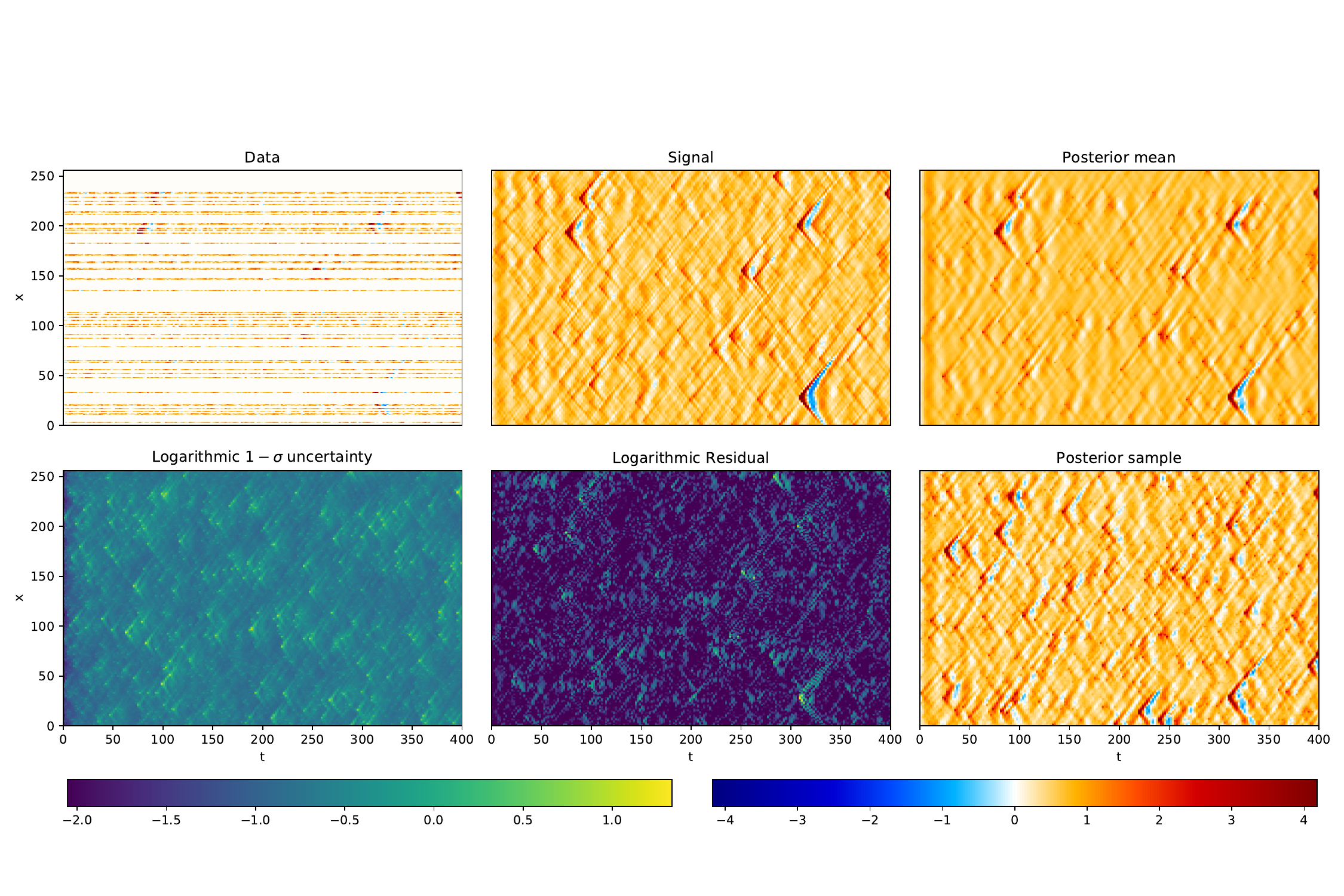}
	\centering
	\caption{{\bf Top:} The left panel shows the sparse and noisy measurement data, drawn from the synthetic signal (middle panel) and the corresponding reconstruction (right panel). {\bf Bottom:} Natural logarithmic one-sigma posterior uncertainty (left panel), natural logarithmic residual of the true signal and the corresponding posterior mean (middle panel), as well as an approximate posterior sample (right panel).}\label{fig:2dsparsesignal}
\end{figure*}

In our last example we aim to perform source detection in the excitation field, in a case where also the dynamic response is unknown. To demonstrate this scenario we again generate a 1+1-dimensional synthetic example where in this case we assume that we are only able to measure the temporal evolution of the system at several locations. In particular we measure the temporal evolution at 50 randomly selected locations of the space under consideration. This results in $\approx 80 \%$ of the discrete space-time being unobserved, as the resolution is the same as in the previous example. As before, we assume that the measurements are subject to additive Gaussian noise with $\sigma_n = 0.3$ and also assume the system to be at rest at $t = 0$. The resulting data is shown in the top-left panel of figure \ref{fig:2dsparsesignal}. The unknown excitations are inverse gamma distributed to model strong but sparse excitations. We infer those from measurements of the system at multiple locations together with the dynamics encoding function $f$. The hyper-parameters of the prior for $f$ (see Eq.\ \eqref{eq:gausssmoothn}) are set to $(\sigma, \gamma) = (2.9, 1.7)$. The system used to generate the data exhibits damped traveling waves described by
\begin{equation}\label{eq:sparse}
\mathcal{L} = \partial_t^2 - c^2 \partial_x^2 + m^2 + \alpha \partial_t - \beta \partial_x \ ,
\end{equation}
with $(c,m,\alpha,\beta) = (1.2,0.04,-0.013,0.005)$. From an information theoretical point of view this setup is very similar to the previous one since we only changed the measurement response to describe measurements of the temporal evolution at several locations, as well as the prior for excitations to be an inverse gamma prior. Consequently also the inference can be performed in the same way as before.

The setup as well as the reconstruction of the field evolving in space-time are shown in figure \ref{fig:2dsparsesignal}. We see that the reconstruction recovers many sources and the corresponding propagation. The algorithm uses information from the response of strong excitations to reconstruct the Green's function of the system. Due to the assumed homogeneity in space-time, this information helps to improve the overall reconstruction in other regions. The quality of the reconstruction of single excitations additionally depends on the surrounding measurement scenario.

\begin{figure*}[htp]
	\centering
	\includegraphics[scale=0.45, angle=0]{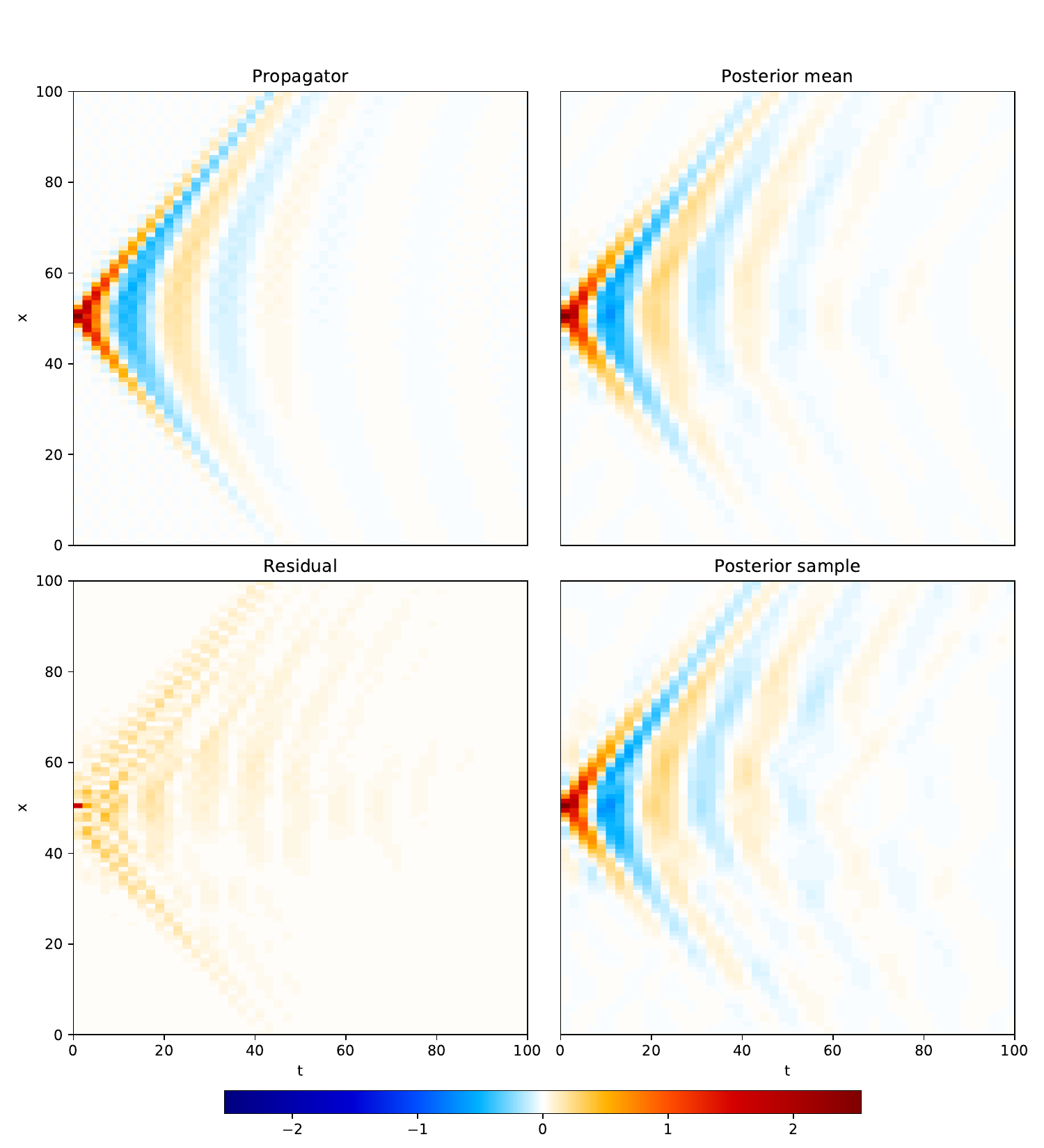}
	\centering
	\caption{{\bf Top:} True Green's function of the process defined in eq.\ \ref{eq:sparse} (left) and corresponding posterior mean (right). {\bf Bottom:} Residual of the true Green's function and the reconstruction (left) and a posterior sample for the Green's function (right).}\label{fig:2dsparseprop}
	\centering
	\includegraphics[scale=0.45, angle=0]{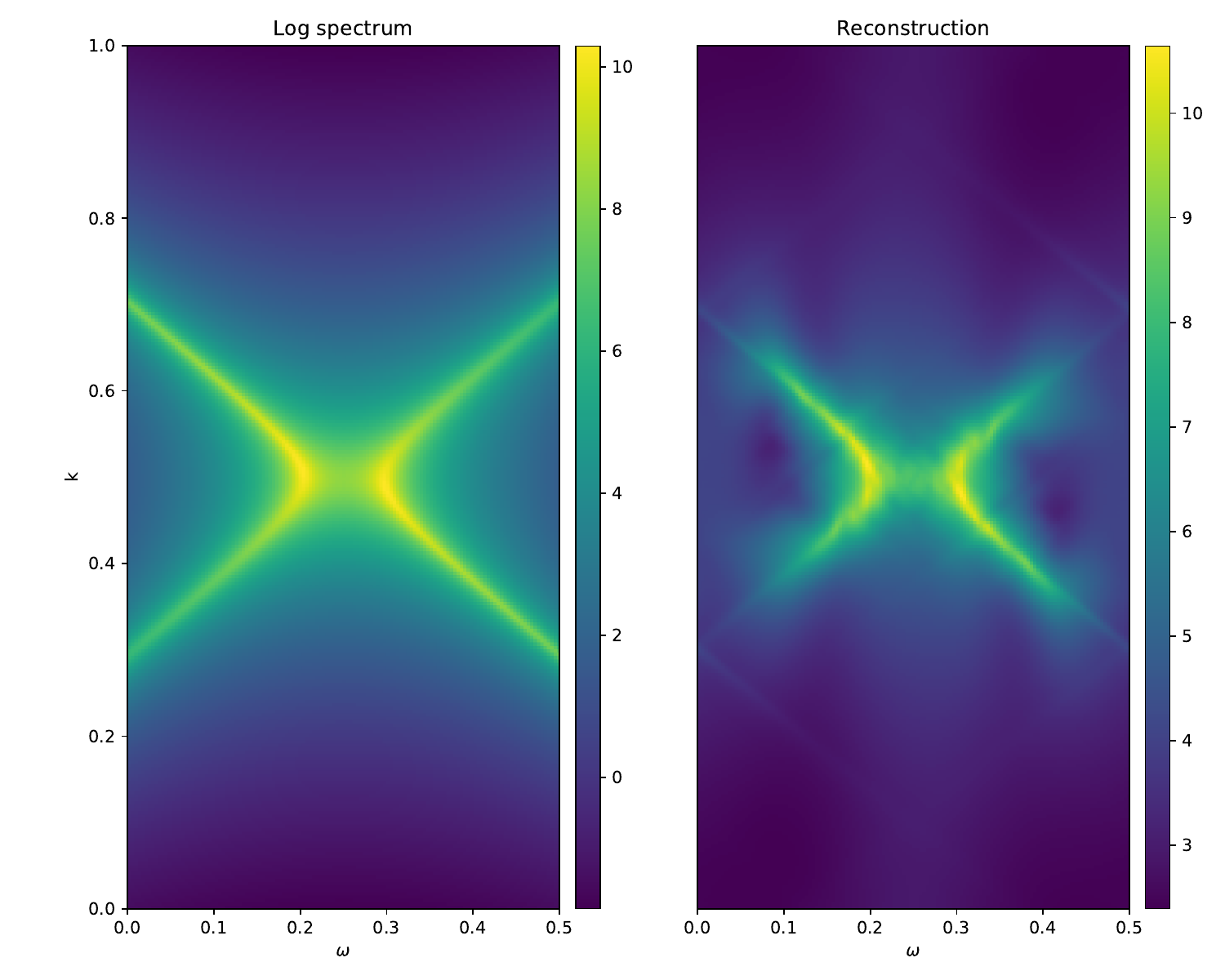}
	\centering
	\caption{Natural logarithmic spectrum of the true Green's function (left) as well as the corresponding posterior mean (right).}\label{fig:2dsparsespec}
\end{figure*}

In figures \ref{fig:2dsparseprop} and \ref{fig:2dsparsespec} we depict the dynamic propagator as well as the spectrum and reconstructions. We can validate that the Green's function was indeed reconstructed correctly, within uncertainties.
We conclude that the task of source detection is possible even in cases where the underlying dynamics is unknown, as long as the assumptions of spatio-temporal homogeneity, causality, and locality hold.

\section{Conclusion}\label{sec:conclusion}
In this work we considered the problem of reconstructing a random field $s$, defined in space and time, together with its correlation structure $S$ from noisy and incomplete data $d$ about $s$. We have shown that this Bayesian hierarchical inference problem can be reformulated to a (theoretically) equivalent problem by means of a generative process, where we aim to infer an excitation field $\xi$ as well as the dynamic response $G$. Ultimately the eigen-spectrum of $G$ was encoded in the dynamics encoding field $f$ and the propagation-speed encoding parameter $\mathbf{C}$. Together with the excitations $\xi$ they denote the quantity of interest $a$ of the inference problem. We proposed a Gaussian process prior for $f$ which gives rise to a non-parametric description of the dynamic response $G$. This gives rise to a non-linear generative process for $s$. As the proposed method is also applicable for non-Gaussian prior distributions for $\xi$ it can also model a variety of other, physically motivated, prior distributions for $s$. This flexibility is discussed for the example of an inverse gamma distribution for the excitations.

To demonstrate its applicability, the proposed method is applied to several synthetic data examples. These include a one dimensional example where the excitations were known and only $G$ had to be inferred, a 1+1 dimensional example with unknown Gaussian distributed excitations as well as an example with inverse gamma distributed $\xi$ and a measurement response that is sparse in the spatial domain.

As we restricted the prior assumptions for $G$ to the physically motivated concepts of space-time homogeneity, locality, and causality, the method appears to be applicable in a wide range of problems. One particular strength is the non-parametric formulation of the Green's function. This becomes important in scenarios where physical models cannot provide a simple parametric description of evolution so far, to describe the Green's function. In addition, a probabilistic description of excitations is sufficient for inference. Consequently the method is still applicable in cases where external influences cannot be described completely.

All in all, we believe that the proposed method is capable of dealing with current as well as upcoming inference problems involving fields defined over space and time, arising from the context of astrophysical imaging.

\appendix
\section{Light cone prior on a discretized space}\label{ap:cone}
As discussed in section \ref{sec:prior} the concept of causality in space-time results in a restriction of propagation within a light cone. In a continuous description, this restriction is realized via a convolution with a step function of the form
\begin{equation}\label{ap:thet}
\Theta(l^2) = \Theta\left( t^2 - \mathbf{x}^\dagger \mathbf{C}^{-1} \mathbf{x} \right) \ .
\end{equation}
Due to the fact that our calculations are ultimately performed on a finite grid, this definition appears to be somewhat problematic, as it introduces boundary effects along the edges of the step function, when realized on a discretized space. In addition, if we aim to elevate $\mathbf{C}$ to be an unknown parameter of the problem that has to be inferred, gradient based methods are no longer applicable due to the fact that the gradient is zero almost everywhere in space-time (or not defined on the boundary). Therefore we seek to find a way to relax the sharp boundary introduced via the cone, without loosing its useful properties. To do so we borrow an idea from quantum field theory where it turns out that these sharp boundaries are ``smeared" out when considering them on a quantum scale. In our case the ``quantum" scale can be regarded as the resolution of the discrete representation of space-time, although this analogy is purely artificial.

To achieve this relaxation, consider the following quantity
\begin{equation}
\Delta = \sqrt{- l^2} = \sqrt{- \left(t^2 - \mathbf{x}^\dagger \mathbf{C}^{-1} \mathbf{x} \right) } \ .
\end{equation}
It has two useful properties: For causal (including time-like as well as light-like) points the real part of $\Delta$, $\Re(\Delta)$, is zero as the square root is taken from a negative number. For non-causal (space-like) points the real part becomes positive. Furthermore, for fixed $t$, $\Re(\Delta)$ is asymptotically linear in $x$. Therefore, if we consider a Gaussian in $\Re(\Delta)$,
\begin{equation}
\theta(\Delta) \equiv \exp\left(- \frac{1}{2 \sigma^2} \Re(\Delta)^2\right) \ ,
\end{equation}
we notice that this quantity remains one within the light cone, while it asymptotically falls off like a Gaussian in $x$, for fixed $t$. Here $\sigma$ stands for an optional scaling parameter which controls the width of the Gaussian. In all applications of this paper we replace the function $\Theta$ (Eq.\ \eqref{ap:thet}) with $\theta$ and set $\sigma$ to the size of a few pixels.

\newpage

% Acknowledgements
\medskip
\textbf{Acknowledgements} \par %delete if not applicable))
We would like to thank Jakob Knollm{\"u}ller, Philipp Arras and Margret Westerkamp for fruitful discussions, Martin Reinecke for his contributions to NIFTy, and two anonymous referees for numerous comments that significantly improved the mathematical and overall presentation of the subject.
\medskip

\textbf{Conflict of interest} \par %delete if not applicable))
The authors declare no conflicts of interest.
% References
\medskip

% Use the following code if you wish to generate your bibliography with BibTeX;
% replace the string "MSP-template" below with the name(s) of
% the BibTeX data base(s) you want to use.
% The resulting bibliography-output (the content of the .bbl file)
% must be pasted back into this file before submission.
% Please also include your BibTeX data base file(s) in your submission
% so that we can re-run BibTeX if necessary.
%
\bibliographystyle{MSP}
\bibliography{FDI}

% Figures/tables and captions
% Permission statements are required for all figures reproduced or adapted from previously published articles/sources. Please also ensure that all necessary permissions to reproduce images have been received
% Please remove these statements for original figures

% Table of contents entry should be 50 - 60 words long
% Image should be 55 mm broad and 50 mm high or 110 mm broad and 20 mm high

\begin{figure}
\textbf{Table of Contents}\\
%\tableofcontents
\medskip
  \includegraphics{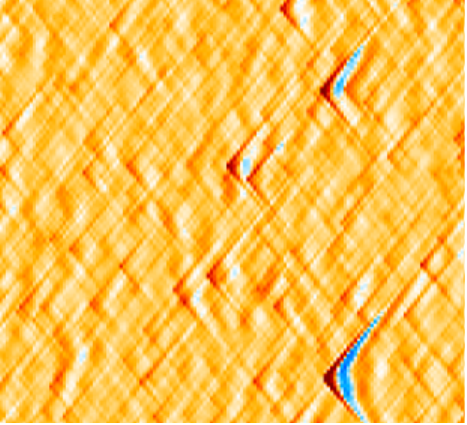}
  \medskip
  \caption*{This work presents a Bayesian inference method to infer space-time homogeneous stochastic processes from noisy and incomplete observations of a single realization of the process. It relies on a reformulation in terms of external random exitations and the dynamic response of the system and performs a joint inference thereof by means of a variational approximation of the posterior.}
\end{figure}

\end{document}